\documentclass[twocolumn]{aastex62}
\pdfoutput=1 
\usepackage{amsmath,amstext}
\usepackage[utf8]{inputenc}
\usepackage{apjfonts} 
\usepackage[figure,figure*]{hypcap}
\usepackage[inline]{enumitem}   
\usepackage{longtable}
\usepackage{tabularx}
\usepackage{enumerate}
\usepackage{graphicx}

\shorttitle{Planet distribution}
\shortauthors{Baron et al.}
\newcommand{\mj}{\ensuremath{M_{\rm Jup}}}

\begin{document}

\title{Constraints on the occurrence and distribution of 1--20 \mj\ companions to stars at separations of 5--5000\,au from a compilation of direct imaging surveys}

\author{Frédérique Baron}
\affiliation{Institut de Recherche sur les Exoplanètes, Département de Physique, Université de Montréal, Montréal, QC H3C 3J7, Canada}

\author{David Lafrenière}
\affiliation{Institut de Recherche sur les Exoplanètes, Département de Physique, Université de Montréal, Montréal, QC H3C 3J7, Canada}
\author{Étienne Artigau}
\affiliation{Institut de Recherche sur les Exoplanètes, Département de Physique, Université de Montréal, Montréal, QC H3C 3J7, Canada}
\author{Jonathan Gagné}
\affiliation{Institut de Recherche sur les Exoplanètes, Département de Physique, Université de Montréal, Montréal, QC H3C 3J7, Canada}
\author{Julien Rameau}
\affiliation{Institut de Recherche sur les Exoplanètes, Département de Physique, Université de Montréal, Montréal, QC H3C 3J7, Canada}
\affiliation{Université de Grenoble Alpes, CNRS, IPAG, F-38000 Grenoble, France}
 \author{Philippe Delorme}
 \affiliation{Université de Grenoble Alpes, CNRS, IPAG, F-38000 Grenoble, France}
\author{Marie-Eve Naud}
\affiliation{Institut de Recherche sur les Exoplanètes, Département de Physique, Université de Montréal, Montréal, QC H3C 3J7, Canada}

\begin{abstract}
We present the first statistical analysis of exoplanet direct imaging surveys combining adaptive optics imaging at small separations with deep seeing-limited observations at large separations allowing us to study the entire orbital separation domain from 5 to 5000~au simultaneously. Our sample of 344 stars includes only confirmed members of nearby young associations and is based on all AO direct-imaging detection limits readily available online, with addition of our own previous seeing limited surveys. Assuming that the companion distribution in mass and semi-major axis follows a power law distribution and adding a dependence on the mass of the host star, such as $d^2n\propto fM^{\alpha}a^{\beta} (M_\star/M_{\odot})^{\gamma}$d$ M $d$a$, we constrain the parameters to obtained $\alpha=-0.18^{+0.77}_{-0.65}$, $\beta=-1.43^{+0.23}_{-0.24}$, and $\gamma=0.62^{+0.56}_{-0.50}$,at a 68\% confidence level, and we obtain $f=0.11^{+0.11}_{-0.05}$, for the overall planet occurrence rate for companions with masses between 1 to 20~\mj\ in the range 5--5000~au. Thus, we find that occurrence of companions is negatively correlated with  semi-major axis and companion mass (marginally) but is positively correlated with the stellar host mass. Our inferred mass distribution is in good agreement with other distributions found previously from direct imaging surveys  for planets and brown dwarfs, but is shallower as a function of mass than the distributions inferred by radial velocity surveys of gas giants in the 1--3\,au range.  This may suggest that planets at these wide and very-wide separations represent the low-mass tail of the brown dwarfs and stellar companion distribution rather than an extension of the distribution of the inner planets.

\end{abstract}


\section{Introduction}
Over the last 15 years, many teams -- using several telescopes on the ground and in space-- surveyed young nearby stars to uncover new planets with direct imaging \citep{marois_direct_2008,lafreniere_gemini_2007,delorme_direct-imaging_2013,macintosh_discovery_2015,naud_psym-wide:_2017}.
All in all, about 1000 unique stars were observed in search of planets using first-generation adaptive optics (AO) systems, seeing-limited imaging, or space-based telescopes. The orbital separations probed by these surveys range from several au to hundreds and even thousands of au, while the detectable planet masses are restricted to that of Jupiter or higher. 
Some of these surveys targeted only higher mass stars \citep{vigan_international_2012,nielsen_gemini_2013,rameau_survey_2013}, others aimed only at low mass stars \citep{bowler_planets_2015,lannier_massive:_2016,galicher_international_2016,naud_psym-wide:_2017}, and some surveyed stars of all spectral types \citep{lafreniere_gemini_2007,heinze_constraints_2010,biller_gemini/nici_2013,chauvin_vlt/naco_2015,meshkat_direct_2017,uyama_seeds_2017,baron_weird:_2018,stone_leech_2018}.
Although only a few planets have been found through these efforts, the resulting large data set can be used to investigate the occurrence rate and distribution of planets as well as their dependence on the star properties.

Such studies are necessary to gain knowledge about the formation and evolution of planets at the large orbital separations probed by direct imaging. This is particularly important since the standard planet formation models --core accretion or disk instability-- struggle to form planets beyond 100\,au.

One of the first attempts to constrain the orbital separation distribution and the mass distribution of Jupiter-like planets was made by \cite{tabachnik_maximum-likelihood_2002}. Using data on 72 planets found by radial velocity (RV) and a distribution such as $dn=C(M)^{-\alpha}(P)^{-\beta} \textrm{d}lnM \textrm{d}lnP$, where $M$ is the planet mass and $P$ the orbital period, they inferred that $\alpha=0.11\pm0.10$ and $\beta=-0.27\pm0.06$ for $M\lesssim10$~\mj\ and 2~d $<P<$ 10~yr. This idea was then pushed further by \cite{cumming_keck_2008} who also used a power law to fit the distribution of planets with masses over 0.3~\mj\ and periods less than 2000 days detected using radial velocity measurements of FGK stars. With a mass-period distribution of $\textrm{d}^2n=C(M)^{\alpha}(P)^{\beta} \textrm{d}lnM \textrm{d}lnP$, they obtained a constraint on the parameters of $\alpha=-0.31\pm0.20$ and $\beta=0.26\pm0.10$. Also based on radial velocity measurements of 166 stars, \cite{howard_occurrence_2010} found that planet occurrence increases with decreasing planet mass, such as  $\textrm{d}n=0.39M^{-0.48} \textrm{d}lnM$. \cite{dong_fast_2013} studied Kepler planets with P $<$ 250 days and found that the distribution of planets in terms of periods, $\textrm{d}n/\textrm{d}lnP$, is proportional to $P^{0.7\pm0.1}$ for Neptune-sized planet and agrees with a flat distribution for Super-Earth or Earth-sized planets.
\cite{fernandes_hints_2018}, based on transit and radial velocity data, described the distribution of giant planets as a broken power law in semi-major axis, showing initially an increase of planet occurrence with semi-major axis, a turnover at about 3~au, followed by a decrease. They also found a power law distribution in planet mass, showing an increase in occurence for lower masses.
All of those studies mostly focus on close-in planets of various masses, and there are very few constraints on semi-major axes greater than 10\,au and even fewer over 100\,au. However, the distribution of planets as presented in \cite{cumming_keck_2008} was widely used, and still is, when planning surveys with direct imaging, as a way to predict the planet yield of the survey. Constraining the planet distribution of massive planets on wide orbits is needed to get more accurate planets yields. 

A few constraints on the planet distribution do exist for separations beyond 10\,au from direct imaging data. For instance, \cite{heinze_constraints_2010} used their adaptive optics planet imaging survey to rule out with 90\% confidence a distribution as in \cite{cumming_keck_2008} at separations up to 110\,au. \cite{brandt_statistical_2014} used a combined sample of direct imaging data to model the population of companion with masses 5--70~\mj\ and semi-major axes of 10--100\,au with a power law. They found that $dn \propto M^{-0.65\pm0.60} a^{-0.85\pm0.39}$, which does not agree with the distribution of planets from \cite{cumming_keck_2008} and hints that the low-mass companions in their sample represent the low-mass tail of the brown dwarfs distribution.
\cite{reggiani_vlt/naco_2016} showed that the results from direct imaging surveys searching for substellar companions around Sun-like stars are consistent with an extrapolation of the \cite{cumming_keck_2008} distribution to larger separations combined with the log-normal brown dwarf mass distribution from \cite{raghavan_survey_2010}. Lastly, \cite{meyer_m-dwarf_2018} studied planets with masses between 1 and 10 \mj\ and separations between 0.07 and 400\,au and found that the semi-major axis distribution is best described by a log-normal distribution peaking at about 3\,au. For larger separations, there are virtually no constraints to date.


\begin{table*}
\center
\caption{Young Moving Groups\label{asso}}
\begin{tabular}{ccccc}
\hline\hline Name& Short Name &Distance& Age & Ref. \\ &(pc)&(Myr)&\\\hline
$\beta$ Pictoris & BPIC&9-73 & 24$\pm$3 & \cite{shkolnik_all-sky_2017} \\\hline
AB Doradus &ABD& 37-77 & 149$^{+51}_{-19}$ & \cite{bell_isochronal_2016}\\\hline
Argus &ARG&  29-118 & 40-50 & \cite{torres_young_2008,zuckerman_nearby_2018}\\\hline
Carina &CAR& 46-88 & 45$^{+11}_{-7}$ &\cite{bell_isochronal_2016} \\\hline
Columba &COL&  35-81 & $42^{+6}_{-4}$&\cite{bell_isochronal_2016}  \\\hline
Tucana-Horologium&THA& 36-71 & 45$\pm$3 & \cite{bell_isochronal_2016} \\\hline
TW Hya&TW&8-92 & 10$\pm$3 & \cite{bell_isochronal_2016}\\\hline
Hercules-Lyra&HLY& $\sim 30$  &  260$\pm50$& \cite{eisenbeiss_hercules-lyra_2013}\\\hline
Lower Centaurus Crux&LCC& $\sim 140$ & $16\pm2$&\cite{pecaut_star-formation_2016} \\\hline
$\epsilon$ Chamaeleontis&EPSC  & $\sim 100 $ & $3.7\pm4.6$& \cite{murphy_new_2015}\\\hline
Upper Scorpius &US& $\sim130 $ & $10\pm2$&\cite{pecaut_star-formation_2016} \\\hline
Octans&OCT& $\sim 130$ & $35\pm5$&\cite{murphy_re-examining_2013}\\\hline
Carina-Near&CN& $\sim 30  $ & $\sim 200$&\cite{zuckerman_carina-near_2006}\\\hline\hline
\end{tabular}
\end{table*}

In this work, we merged different direct imaging surveys to constrain the occurrence rate and distribution of companions with masses between 1 and 20 ~\mj\ at orbital separations of 5 to 5000\,au. Section~\ref{sample} describes the sample of stars and observations taken from the online archive DIVA and some other previous surveys made by our team.
In Section~\ref{analysis}, we first establish a planet detection completeness map for each target in our sample, and then use those to determine the occurrence rate of companions and, through a Markov chain Monte Carlo (MCMC) approach, constrain the parameters governing their distributions. We discuss our results and their implications in Section~\ref{discussion} and conclude in Section~\ref{conclu}.

\section{Sample}\label{sample}

We assembled a sample of stars that were observed by AO direct-imaging planet searches, as well as by the seeing-limited PSYM-WIDE \citep{naud_psym-wide:_2017} and WEIRD \citep{baron_weird:_2018} surveys. We focused on the stars that are confirmed members of young moving groups with ages of less than 300~Myr, meaning that they have a radial velocity measurement, a trigonometric parallax and $XYZUVW$ values consistent with the moving group spatial's position and space velocity, as well as independent signatures of youth, such as spectroscopic signs of low-gravity, strong X-ray or UV emission or lithium absorption. Table~\ref{asso} presents the young moving groups to which our stars belong, namely TW Hya \citep{de_la_reza_discovery_1989,kastner_x-ray_1997}, $\beta$ Pictoris \citep{zuckerman__2001},
AB Doradus \citep{zuckerman_ab_2004}, Tucana-Horologium \citep{torres_new_2000,zuckerman_tucana_2001}, Carina \citep{torres_young_2008}, Columba \citep{torres_young_2008}, Argus \citep{makarov_moving_2000}, Carina-Near \citep{zuckerman_carina-near_2006}, Upper Scorpius \citep[USCO,][]{pecaut_star-formation_2016}, Lower Centaurus Crux  \citep[LCC,][]{pecaut_star-formation_2016}, $\epsilon$ Chamaeleontis \citep{murphy_re-examining_2013}, Hercules-Lyra \citep{eisenbeiss_hercules-lyra_2013} or Octans \citep{murphy_re-examining_2013}. We have included the star that is a member of Hercules-Lyra, even if \cite{mamajek_pre-gaia_2015} indicated that Hercules-Lyra might be a stream and not a real association. 

The WEIRD survey \citep{baron_weird:_2018} surveyed 177 stars of all spectral types using deep seeing limited imaging to search for giants planets on very wide orbits. A typical completeness of 2\,\mj\ is reached, while some stars of the younger/nearer groups have a 1\,\mj\ detection limit, at separations between 1000 and 5000\,au. We added all the objects from the WEIRD survey in our sample, as they are all bona fide members of young associations. 

The PSYM-WIDE survey \citep{naud_psym-wide:_2017} observed 95 M dwarfs using seeing-limited imaging, out of which only 10 were bona fide members of nearby young moving groups, the others being non-confirmed candidate members at the time of publication. However, using the \textit{Gaia} DR2 release \citep{lindegren_gaia_2018, gaia_collaboration_gaia_2018} and the web tool BANYAN $\Sigma$ from \cite{gagne_banyan._2018}, we confirmed the membership of 34 stars out of those 85 candidates; see Table~\ref{bfpsym}. The total number of bona fide members from PSYM-WIDE used in our study is thus 44. The survey reached good completeness for a mass of 10 \mj\ or more at semi-major axes larger than 1000\,au. 

To complement the above seeing-limited observations, sensitive to the widest orbital separations, we used the DIVA archive \citep{vigan_vlt/naco_2017} to extract data for 119 stars that are bona fide members of young associations of less than 300\,Myr and that were observed at closer separations by \cite{masciadri_search_2005}, \cite{biller_imaging_2007}, \cite{lafreniere_gemini_2007}, \cite{kasper_novel_2007}, \cite{chauvin_deep_2010}, \cite{heinze_constraints_2010}, \cite{vigan_international_2012}, \cite{rameau_survey_2013}, \cite{chauvin_vlt/naco_2015}, \cite{meshkat_searching_2015}, or \cite{meshkat_searching_2015-1}. Out of the 119 stars, 73 have also been observed by the WEIRD or the PSYM-WIDE surveys at larger separations. Overall, the completeness maps of these targets reach good completeness at 3~\mj\ over a range of semi-major axes of 50 to 5000\,au. 
 
We also used data from the AO survey of Upper Scorpius stars of \cite{lafreniere_adaptive_2014}. They list 91 stellar members of Upper Scorpius, and 84 of them have a parallax in \textit{Gaia} DR2. One of them, HIP~78265, was rejected from our sample because its new \textit{Gaia} parallax puts it at a distance of 590\,pc, which is too far from the other members of Upper Scorpius. All other stars with parallaxes from that study were kept. This survey probes an intermediate range of semi-major axes compared to the above AO imaging survey, with a good completeness between 200 and 800\,au for companions with masses as low as 10\,\mj, as the members of this association are further away than most other targets in the sample. 
 
\begin{figure*}[t!]
\centering
\includegraphics[width=\linewidth]{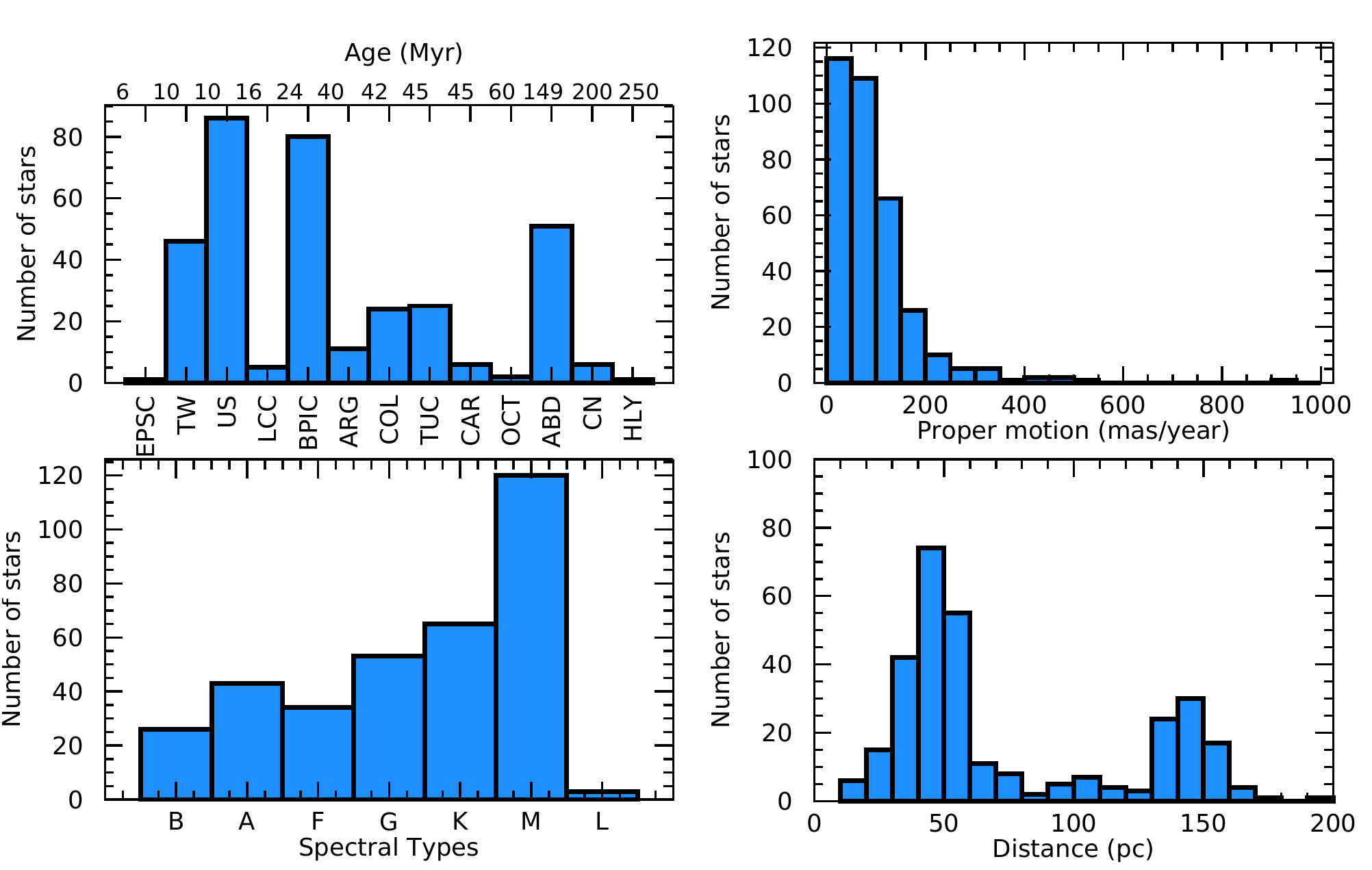}
\caption{\label{histo} Histograms of the number of stars in each association, distances (pc), proper motions (mas) and spectral types of the targets in the sample.}
\end{figure*}

Table~\ref{listecomplete} lists the 344 unique stars in our sample, along with their right ascension, declination, spectral type, proper motion in right ascension and declination, association, and distance. Figure~\ref{histo} presents the summary of our sample. The median target has a distance of 50~pc, a proper motion of 80\,mas yr$^{-1}$ and an age of 24\,Myr. The spectral types of the targets range from B to L dwarfs, and most of the targets are M dwarfs.

\startlongtable
\begin{deluxetable*}{ccccc}
\tablecaption{Confirmed members from PSYM-WIDE\label{bfpsym}}
\tablewidth{700pt}
\tabletypesize{\scriptsize}
\tablehead{
\colhead{2MASS Name} & \colhead{RA} &
\colhead{DEC} & \colhead{Probability} &
\colhead{Association} \\
\colhead{} & \colhead{(J2000.0)} & \colhead{(J2000.0)} & \colhead{\%} &
\colhead{} 
}
\startdata
        J00325584-4405058   &                  8.2326770   &                 -44.084965   &                       99.3   &                        ABD  \\
        J00374306-5846229   &                  9.4294440   &                 -58.773033   &                       99.7   &                        THA  \\
        J01123504+1703557   &                  18.146006   &                  17.065475   &                       99.1   &                        ABD  \\
        J01521830-5950168   &                  28.076262   &                 -59.838001   &                    $>$99.9   &                        THA  \\
        J02045317-5346162   &                  31.221569   &                 -53.771183   &                    $>$99.9   &                        THA  \\
        J02070176-4406380   &                  31.758289   &                 -44.112339   &                    $>$99.9   &                        THA  \\
        J02215494-5412054   &                  35.478949   &                 -54.201511   &                    $>$99.9   &                        THA  \\
        J02224418-6022476   &                  35.684107   &                 -60.379906   &                       80.0   &                        CAR  \\
        J02340093-6442068   &                  38.503875   &                 -64.701912   &                    $>$99.9   &                        THA  \\
        J02485260-3404246   &                  42.219191   &                 -34.073517   &                       99.8   &                        COL  \\
        J02564708-6343027   &                  44.196205   &                 -63.717438   &                       90.4   &                        CAR  \\
        J03350208+2342356   &                  53.758697   &                  23.709892   &                       99.0   &                       BPIC  \\
        J04091413-4008019   &                  62.308892   &                 -40.133862   &                    $>$99.9   &                        COL  \\
        J04213904-7233562   &                  65.412690   &                 -72.565613   &                    $>$99.9   &                        THA  \\
        J04363294-7851021   &                  69.137280   &                 -78.850594   &                       96.0   &                        ABD  \\
        J04402325-0530082   &                  70.096891   &                 -5.5022970   &                       96.4   &                         CN  \\
        J04440099-6624036   &                  71.004021   &                 -66.402084   &                       97.1   &                        THA  \\
        J04571728-0621564   &                  74.322039   &                 -6.3656870   &                       99.8   &                        ABD  \\
        J05241317-2104427   &                  81.054884   &                 -21.078550   &                    $>$99.9   &                        COL  \\
        J05335981-0221325   &                  83.499224   &                 -2.3590290   &                    $>$99.9   &                       BPIC  \\
        J05395494-1307598   &                  84.978924   &                 -13.133292   &                       95.4   &                        COL  \\
        J06112997-7213388   &                  92.874897   &                 -72.227448   &                       94.8   &                        CAR  \\
        J08173943-8243298   &                  124.41432   &                 -82.724945   &                       99.7   &                       BPIC  \\
        J12383713-2703348   &                  189.65473   &                 -27.059681   &                       99.7   &                        ABD  \\
        J18420694-5554254   &                  280.52895   &                 -55.907082   &                       99.8   &                       BPIC  \\
        J19560294-3207186   &                  299.01626   &                 -32.127125   &                       93.5   &                       BPIC  \\
        J20004841-7523070   &                  300.20174   &                 -75.385284   &                       99.8   &                       BPIC  \\
        J21100535-1919573   &                  317.52232   &                 -19.332603   &                       99.8   &                       BPIC  \\
        J22021626-4210329   &                  330.56775   &                 -42.175831   &                       99.4   &                        THA  \\
        J23131671-4933154   &                  348.31962   &                 -49.554298   &                    $>$99.9   &                        THA  \\
        J23285763-6802338   &                  352.24016   &                 -68.042747   &                       98.6   &                        THA  \\
        J23320018-3917368   &                  353.00077   &                 -39.293564   &                    $>$99.9   &                        ABD  \\
        J23452225-7126505   &                  356.34272   &                 -71.447380   &                    $>$99.9   &                        THA  \\
        J23474694-6517249   &                  356.94561   &                 -65.290260   &                    $>$99.9   &                        THA  \\
\enddata
\end{deluxetable*}
%


Seven stars in our sample are known hosts of companions ($M<20$~\mj) on wide orbits : 51 Eri, HR~8799, $\beta$~Pictoris, AB~Pic, Gu~Psc, TWA~27 and 1RXS~J160929.1-210524. Each system is described briefly below. 

\begin{itemize}

\item 51 Eri\,b is a 2--10~\mj\ planet orbiting the F0IV star 51~Eri at a projected separation of $\sim$14\,au; it was found with GPI at Gemini \citep{macintosh_discovery_2015}. While the star is part of our sample, the companion was not detected in the data we compiled.

\item HR~8799 hosts four planets of 7$^{+4}_{-2}$, 10${\pm3}$, 10${\pm3}$ and 9${\pm4}$ \mj\ \citep{marois_direct_2008,marois_images_2010} at semi-major axes of, respectively, $70^{+0.19}_{-0.18}$, $43.1^{+1.3}_{-1.4}$, $26.2^{+0.9}_{-0.7}$ and $16.2\pm0.5$\,au, assuming stable coplanar orbits \citep{wang_dynamical_2018}.  \cite{rameau_survey_2013}, whose data are part of our study, were able to recover planets b, c and d, but not planet e due to the small parallactic angle rotation of the observation. We assume in this work that HR~8799 is an A5V star member of the Columba association, although its membership has been questioned. However, signs of youth are present in the planets' spectra and \cite{zuckerman_tucana/horologium_2011} have shown that HR~8799 is younger than the Pleiades. 

\item $\beta$ Pictoris, an A6V star member of the $\beta$ Pictoris moving group, hosts a $11\pm2$\,\mj\ planet \citep{lagrange_probable_2009,snellen_mass_2018} with a semi-major axis of $9.2^{+1.5}_{-0.4}$\,au \citep{millar-blanchaer_beta_2015}.  \cite{rameau_survey_2013} were also able to recover this companion.

\item AB~Pic~b is a $13.5\pm0.5$~\mj\ object at a projected separation of 250\,au from AB~Pic, a K2V star member of the Tucana-Horologium association \citep{chauvin_companion_2005}. It was recovered by \cite{rameau_discovery_2013} and also tentatively recovered by \cite{biller_imaging_2007}.

\item GU~Psc, an M3V star member of the AB Doradus association, is host to a $11\pm2$~\mj\ companion \citep{naud_discovery_2014} at an angular separation of $42\arcsec$. We used the new parallax from \textit{Gaia} DR2 to revise the projected separation estimate to $1998\pm6$\,au. The discovery observations were part of the PSYM-WIDE survey.

\item TWA~27 hosts a 4$\pm$1~\mj\ companion \citep{chauvin_giant_2004} at a projected separation of 46 $^{+37}_{-15}$\,au \citep{blunt_orbits_2017}. The host is a young brown dwarf, member of the TW Hydrae association at 52\,pc. The planet was discovered using VTL/NACO but it was not detected in the images used in our survey. 

\item  As part of the Upper Scorpius survey used in the present study, \cite{lafreniere_direct_2008,lafreniere_directly_2010} found a directly imaged planet around 1RXS~J160929.1-210524, an M0 star member of the Upper Scorpius association. 1RXS~J160929.1-210524b has a mass between 7 and 12~\mj\ \citep{lachapelle_characterization_2015} and a projected separation of 320$\pm$40\,au.

\end{itemize}

The sample of our study thus includes five detected planet-hosting stars and seven detected planets. Four of those orbit BA stars, two orbit FGK stars and one is around an M dwarf. The small number of planets around M dwarfs in our sample may seem surprising, given the large number of M dwarfs in our sample and the relatively large number of companions found by direct imaging around M dwarfs \citep[e.g., ][]{goldman_new_2010,naud_discovery_2014,kraus_three_2014,luhman_discovery_2006,ireland_two_2011,deacon_nearby_2016,dupuy_hawaii_2018,delorme_brown_2013,bowler_planets_2013,artigau_banyan._2015,todorov_discovery_2010,gauza_discovery_2015,itoh_young_2005,luhman_discovery_2009,rebolo_discovery_1998,reid_lp_2006}, but we point out that out of all those detection but only two were found around M bona fide members of young associations \citep{chauvin_giant_2004,naud_psym-wide:_2017}.

\section{Analysis}\label{analysis}

We used the 7-$\sigma$ detection limits as a function of angular separation provided by the DIVA archive, by \cite{baron_weird:_2018}, by \cite{naud_psym-wide:_2017} and by \cite{lafreniere_adaptive_2014} to build the completeness maps for each target. We first defined a 100x100 grid of masses and semi-major axes, with the masses equally spaced in logarithmic scale between 1 and 20~\mj\ and the semi-major axes equally spaced in logarithmic scale between 5 and 5000\,au. At each point of the grid, $10^4$ planets were simulated, each having an eccentricity taken randomly from the beta function eccentricity distribution reported in \cite{kipping_parametrizing_2013}, which is taken from the eccentricity distribution of RV planets as well as a random inclination and orbital phase, which then yield a projected separation for each planet. Following the method described in \cite{baron_weird:_2018} and using the AMES.Cond evolution models \citep{allard_limiting_2001,baraffe_evolutionary_2003} to convert the planets masses to flux, we compared the planet's magnitudes to the detection
limits found earlier to assess the detectability of each planet. If the detection limits were provided in planet-to-star contrast, they were converted into detection limit in planet absolute magnitude using the star's magnitudes. We then obtained completeness maps for all the stars in the sample.
If a given star was observed by two or more surveys, the most sensitive detection probability was adopted at each point of the mass--separation grid, and we assumed that the exoplanet did not move on its orbit between the different observations. As some of the calculations that follow will need it, a similar completeness map was calculated for each star but this time directly over a grid of projected separations instead of semi-major axes.  For this latter approach, there was no need to draw orbital parameters randomly as the fiducial planets were directly generated at projected separations that can be compared directly with the detection limits.

At 5000\,au, about half of the stars in the sample have a $>$~60\% probability of detecting a companion of mass anywhere in the range 1-20~\mj, while about 20\% of the stars have this same probability at 20\,au, and 10\% at 5 au. We choose a lower mass limit of 1~\mj\ as observations from the WEIRD sample reach sufficient completeness ($>$~50\%) at this mass for large semi-major axis ($>1000$~au). The upper mass limit of 20~\mj\ was chosen to exclude the brown dwarf companion population, as radial velocity data suggest a natural dividing line between planets and brown dwarfs somewhere in the 25--45~\mj\ range \citep{sahlmann_possible_2010}.

\begin{figure*}[h] 
\centering
\includegraphics[width=\linewidth]{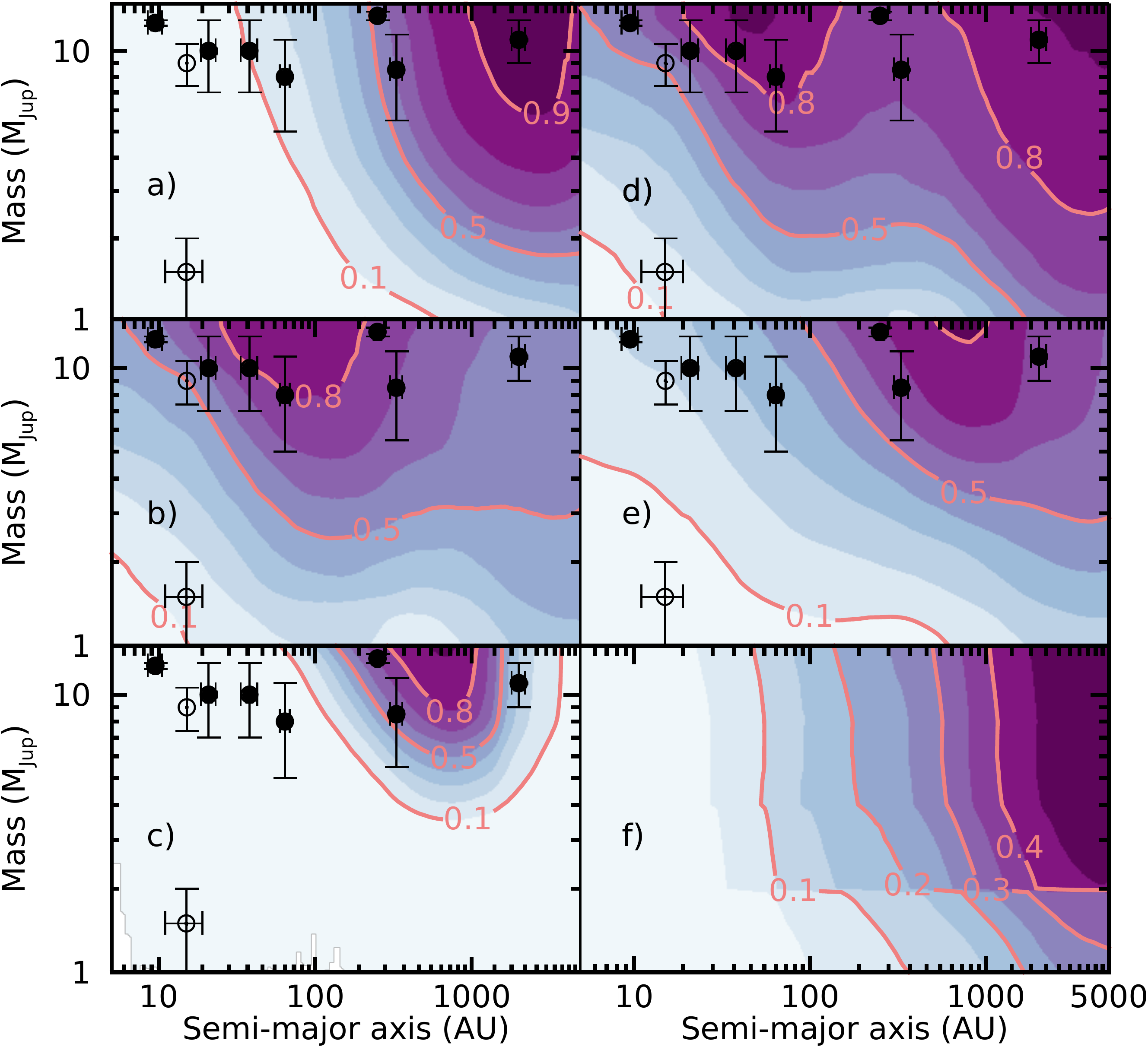}
\caption{\label{maps} Average detection completeness maps of the masses versus the semi-major axis. Filled circles show the known companions detected in the observations used in the present study, while open circles show known companions of stars in our sample that were detected by other surveys. The maps show the average probability of detecting a companion with a mass between 1 and 20~\mj\ as a function of the separation from the host star. Panel (a) is the average completeness map for the seeing-limited observations (WEIRD and PSYM-WIDE) only, sensitive to larger separations, (b) is for AO observations of nearby young associations, sensitive to shorter orbits, (c) is for AO observations of stars the more distance Upper-Scorpius association, sensitive to intermediate separations (d) is for the subset of stars in nearby associations that were observed with both seeing-limited observations and AO, e) is average completeness map for the overall survey and f) is the same as e) using cold-start models. 
}
 \end{figure*}

Figure~\ref{maps} shows the average completeness maps for (a) the 220 objects with WEIRD or PSYM-Wide images (seeing-limited), (b) the 119 targets with only the AO observations from the DIVA archives, (c) the 83 Upper Scorpius targets with only AO observations and (d) the 73 targets that were observed by AO {\em and} either WEIRD or PSYM-WIDE. The known companions discussed above are overplotted. Note that the semi-major axis is used when known; otherwise, the projected separation is used as a semi-major axis. The maps show that the AO images are sensitive to companions with masses of about 7 \mj\ or higher at a completeness of 70\% with a semi-major axis between 50 and 300~au, or a completeness of 50\% for semi-major axes larger than 20~au. The seeing-limited observations, on the other hand, are mostly sensitive to semi-major axes above 500\,au for masses above 3~\mj with a completeness of 70\%. As panel d) demonstrates, combining AO imaging with wide-field imaging enables a good semi-major axis coverage as well as a decent companion mass coverage. Panel e) shows the average completeness maps for the entire survey. Overall, the survey is mostly sensitive to object more massive than 3~\mj\ at semi-major axes between 500 to 1500\,au.

\subsection{Frequency of companions}\label{freq3.1}

Using the individual completeness maps for all targets in the sample and the statistical formalism presented in \cite{lafreniere_gemini_2007}, a frequency $f$ of stars that have at least one companion with a mass between 1 and 20~\mj\ and semi-major axis between 20 and 5000~au was evaluated. For this first analysis, we chose to focus on the 20--5000~au range of semi-major axis, as this is where we reach the most interesting completeness to constrain the occurence rate. Even though we have some sensitivity at smaller separations, extending this analysis to cover smaller separations would lead to larger uncertainties because the completeness is significantly smaller below 20 au.
Assuming that $N$ is the total number of stars in the sample and $k$ is the $i$th star of said sample, then the results of the survey can be summarized by the set $\{d_k\}$, where the value of $d_k$ is 1 if one (or more) companion is detected around star $k$ and $d_k$ is 0 otherwise. If $p_k$ is the probability that such a
star hosts a companion that would be detected given the detection limits of the observations if indeed it was there, then the likelihood of the data for a given value of $f$ is given by the binomial likelihood :

\begin{equation}
\mathcal{L}(\{d_k\}|f) = \prod_{k=0}^{N} (1-fp_k)^{1-d_k}(fp_k)^{d_k} \,.
\end{equation}

If $p(f)$ is the prior probability of $f$, then according to Baye's theorem, the posterior distribution for $f$, in light of the data, is given by, 
\begin{equation}
p(f|\{d_k\}) = \frac{\mathcal{L}(\{d_k\}|f)p(f)}{\int_{0}^{1} \mathcal{L}(\{d_k\}|f)p(f) \textrm{d} f} \,,
\end{equation}

The prior $p(f)$ represents the best knowledge about the posterior distribution of $f$ based only on information that is independent from the current analyses. To apply Bayesian statistics in a
way that only depends on the available data and the given likelihood, it is appropriate to use a non-informative prior \citep{berger_formal_2009}. Here, we used a non-informative Jeffrey's prior, which is appropriate for the binomial likelihood, and is given by :
\begin{equation}
p(f) = \frac{1}{\pi}\frac{1}{\sqrt{f}}\frac{1}{\sqrt{1-f}} \,.
\end{equation}

The maximum of the posterior distribution is obtained for the most likely value for $f$. An equal-tail credible interval ($\alpha$=0.95) can be determined from,

\begin{equation}
\frac{1-\alpha}{2} = \int_{0}^{f_{min}} p(f|{d_{k}})df\,,
\end{equation}

\begin{equation}
\frac{1+\alpha}{2} = \int_{f_{max}}^{1} p(f|{d_{k}})df\,.
\end{equation}

We applied the above procedure to constrain $f$ over various semi-major axis intervals and for planet masses from 1 to 20 Mjup. To compute $p_k$ we simply averaged the above completeness maps over the appropriate region of semi-major axis and planet mass of our grid; this amounts to assuming the planets are distributed uniformly in logarithm scale in both mass and semi-major axis. 

\begin{figure}[h] 
\centering
\includegraphics[width=\linewidth]{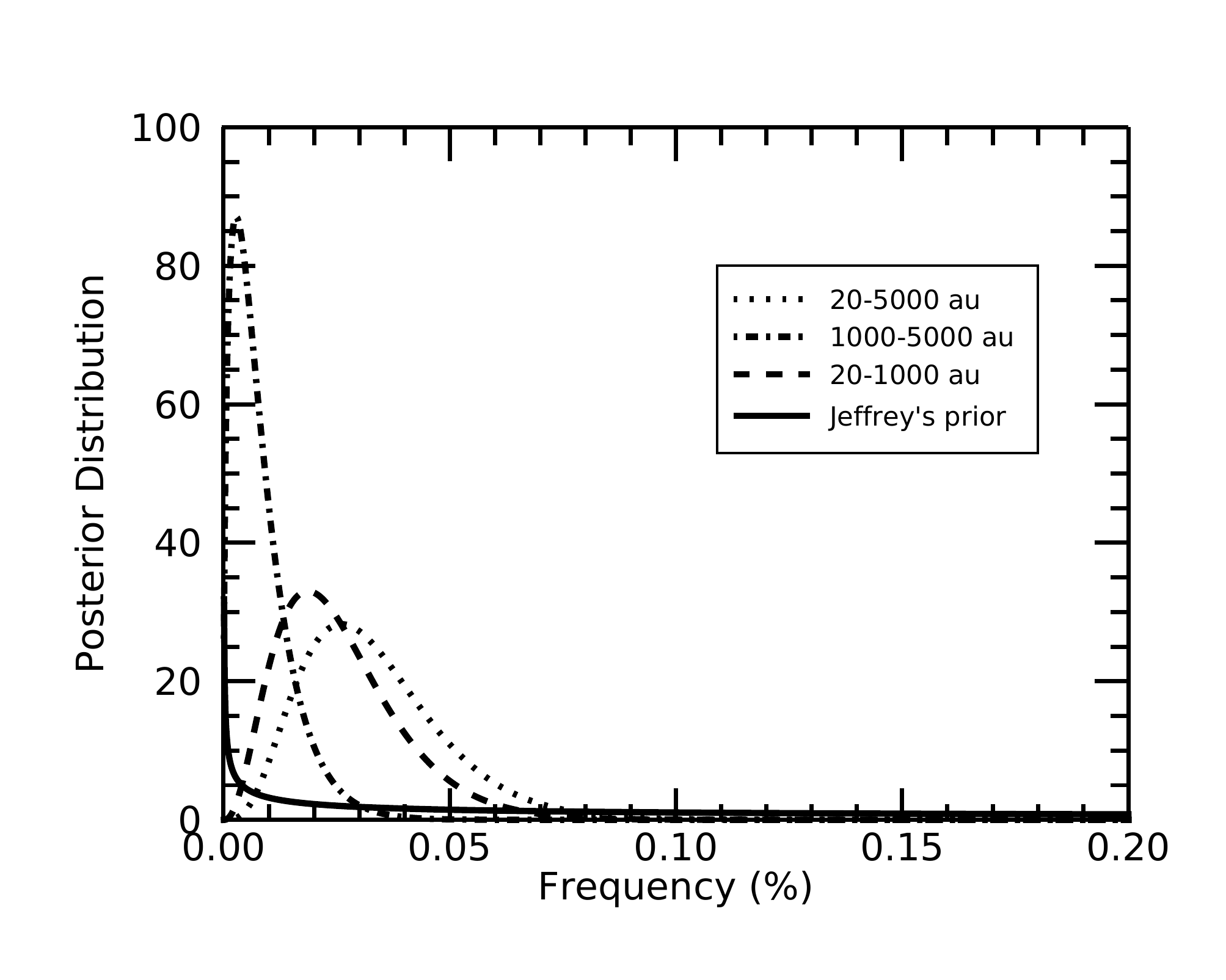}
\caption{\label{postdist} Posterior distributions of the occurence rate of companions of masses between 1 and 20~\mj. The dotted line shows the frequency in the semi-major axis range 20--5000\,au, the dash-dotted line is for the range 1000--5000\,au, and the dashed is for the range 20--1000\,au. The solid line shows the Jeffrey's prior used. 
}
 \end{figure}

Figure~\ref{postdist} shows the posterior distributions obtained for the full semi-major axis range probed by our study, as well as for two sub-ranges, 20--1000~au and 1000--5000~au. 
From these posterior distributions, we can infer a frequency of companions with masses between 1 and 20~\mj\ for the corresponding ranges of semi-major axes. First, for the \hbox{20--1000\,au} range which contains the detection of companions around 4 stars of the sample (AB Pic, HR 8799, HIP 78530 B and 1RXS J160929.1$-$210524), we obtained a frequency of 2.17$^{+6.85}_{-0.73}$\%, at 95\% confidence level. For the 1000--5000~au range which contains only one companion (GU Psc b), we inferred a frequency of 0.3$^{+2.6}_{-0.1}$\%, at a 95\% confidence level. This is much lower than at shorter separations, and is easily understood as this range contains much fewer detections despite having better sensitivities on average. For the overall range of semi-major axes probed here (20--5000~au), we obtained a frequency of 2.61$^{+6.97}_{-1.00}$\%, at 95\% confidence level. 

\begin{figure}[h] 
\centering
\includegraphics[width=\linewidth]{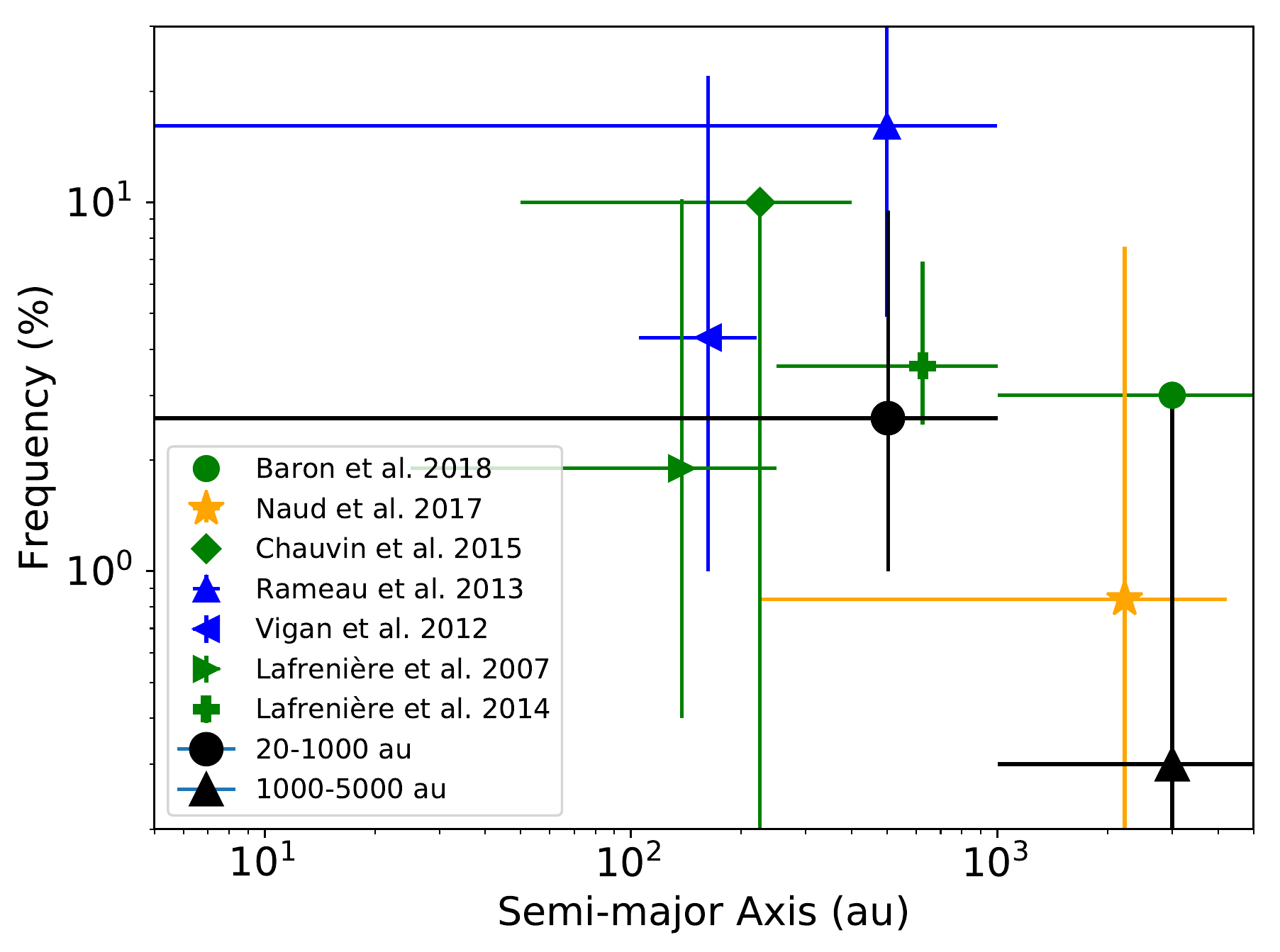}
\caption{\label{freq} Frequency of companions for various ranges of semi-major axes probed by this study and others. The survey that concentrated on M dwarfs is shown in orange, the studies that surveyed A stars in blue and the others are shown in green. The frequency from \cite{baron_weird:_2018} is represented by a circle, the one from \cite{naud_psym-wide:_2017} is drawn as a star, the one from \cite{chauvin_vlt/naco_2015} is a diamond, the one from \cite{rameau_survey_2013} is drawn as a triangle pointing up, the one from \cite{vigan_international_2012} is shown as a left pointing triangle, the one from \cite{lafreniere_gemini_2007} is drawn as a right pointing triangle, and the one for \cite{lafreniere_adaptive_2014} is represented by a thick plus sign.
The frequency from the current analysis is shown as a black square for the range 20--5000\,au, a black triangle for the range 1000--5000\,au, and a dark circle for the range 20--1000\,au.
The horizontal bars represent the ranges of semi-major axes, while the vertical error bars show 95\% credible interval for the companion frequency. 
}
 \end{figure}

Figure~\ref{freq} shows the frequency that we found and those obtained by the individual surveys included in our study over a similar companion mass range and for various ranges of semi-major axes. Here, the horizontal bars represent the ranges of semi-major axes while the vertical error bars correspond to the uncertainty interval on the frequencies (at a 95\% confidence level). The surveys included in our sample cover a wide range of spectral types, but some of the other surveys focused on M dwarfs and others on A stars.  The surveys in our study that focused on wide orbits \hbox{(20--1000\,au)} found overall marginally higher frequencies than those that focused on very wide orbits \hbox{(1000--5000\,au)}. This is consistent with the frequencies that we calculated in both intervals.  Table 3 presents, for context, a compilation of several literature results for the occurrence of giant planets based on direct imaging surveys. We can also compare our results directly to others surveys to show that we get similar results. First we compare to the meta-analysis from \cite{bowler_imaging_2016}. Using the same analysis as previously in the interval of 5--13 \mj\ and 30--300 \,au for all spectral types, we obtain a occurrence of $1.83_{-0.62}^{+5.76}\%$, comparable to the overall occurrence rate of \cite{bowler_imaging_2016} of $0.6_{-0.7}^{+0.5}\%$. We also compare our analysis to studies that targeted M dwarfs, as our survey has a good number of them. In the range 500-5000\,au and 1-13~\mj, \cite{naud_psym-wide:_2017} obtained a frequency of $0.84_{-0.66}^{+6.73}\%$. For the same range of semi-major axis and masses, we get $0.3_{-0.06}^{+2.75}\%$, which is comparable within uncertainties to \cite{naud_psym-wide:_2017}. We also compare with \cite{galicher_international_2016} and we obtained an occurrence of $1.79_{-0.49}^{+7.5}\%$, comparable to their $1.05_{-0.70}^{+2.80}\%$, in the range of 1--14 \mj\ and 20--300\,au.

 \begin{table*}
\center
\caption{Occurrence of giant planets from the litterature \label{chap1freq}}
\begin{tabular}{cccccc}
\hline\hline
Reference& No. of stars& SpT& $f$ & Sep & Mass \\ 
&&&(\%)&(UA)&(\mj)\\\hline

&& Wide orbits  \\\hline

\cite{lafreniere_gemini_2007} & 85& All &$<$28&10--24& 0.5--13  \\
							  &   &      &$<$13&25--50&  \\
							  &   &      &$<$9.3&50--250&   \\\hline  
\cite{heinze_constraints_2010-1} & 54& FGK &$<50$&30--94& 5--13\\\hline 
& &  &$<25$&25--100& 7--13\\
 & &  &$<15$&15--100& 10--13\\
\cite{janson_high-contrast_2011} & 18& BA &$<85$&<100& $<300$\\\hline   
\cite{vigan_international_2012} & 42& A &$8.7_{-6}^{+19.6}$&5--320& 3--14 \\\hline   
\cite{biller_gemini/nici_2013} & 80& All &$<6$&10--150& 1--20 \\
	 & &  &$<7$&10--50& \\\hline  
\cite{nielsen_gemini_2013} & 70& B-A &$<20$&59--460& 4--13 \\\hline 
\cite{rameau_survey_2013} & 59& A-F &$16.1_{-11.2}^{+26.3}$&1--1000& 1--13 \\\hline
\cite{wahhaj_gemini_2013} & 57& debris disk  &$<24$&$>8$& 9--13 \\
 & &  &$<24$&$>63$& 4--13\\\hline 
\cite{lafreniere_adaptive_2014} & 91& All &$4.0_{-1.2}^{+3.0}$&250--1000& 5--40 \\\hline 
 & &  &$<1.8$&50--250& 5--40 \\\hline 

\cite{bowler_planets_2015} & 122& M &$<10.3$&10--100& 1--13 \\\hline 
\cite{chauvin_vlt/naco_2015} & 86& FGK &$<15$&100--500& 5--13 \\ 
 & 86& FGK &$<10$&50--500& 10--13 \\\hline 
\cite{brandt_statistical_2014} & 248& All &$1.0-3.1$&10--100& 5--70 \\\hline 
\cite{bowler_imaging_2016} & 384& BA &2.8$_{-2.3}^{3.7}$&30--300& 5--13 \\ 
 & & FGK &$<4.1$&& \\
 & & M &$<3.9$&&  \\\hline 
\cite{durkan_high_2016} & 73& All &$<9$&100--1000& 0.5--34 \\\hline 
\cite{galicher_international_2016} & 292& M &$1.05_{-0.70}^{+2.80}$&20--300& 0.5--14 \\\hline 
\cite{lannier_massive:_2016} & 58& M &$2.3_{-0.7}^{+2.9}$&8--400& 2--14 \\\hline  
\cite{meshkat_direct_2017} & 277& debris disk &$6.27_{-2.59}^{+3.49}$&10--100& 5--20 \\\hline
\cite{stone_leech_2018} & 98& All &$\sim$25&5--100& 4--14  \\\hline
&&Very wide orbits  \\\hline
\cite{naud_psym-wide:_2017} & 95& K5-L5 &$0.84_{-0.66}^{+6.73}$&500--5000& 5--13 \\ 
\cite{baron_weird:_2018} & 177& All &$<$3&1000--5000& 1--13  \\\hline

\end{tabular}

\end{table*}

\begin{figure*}[h] 
\centering
\includegraphics[width=\linewidth]{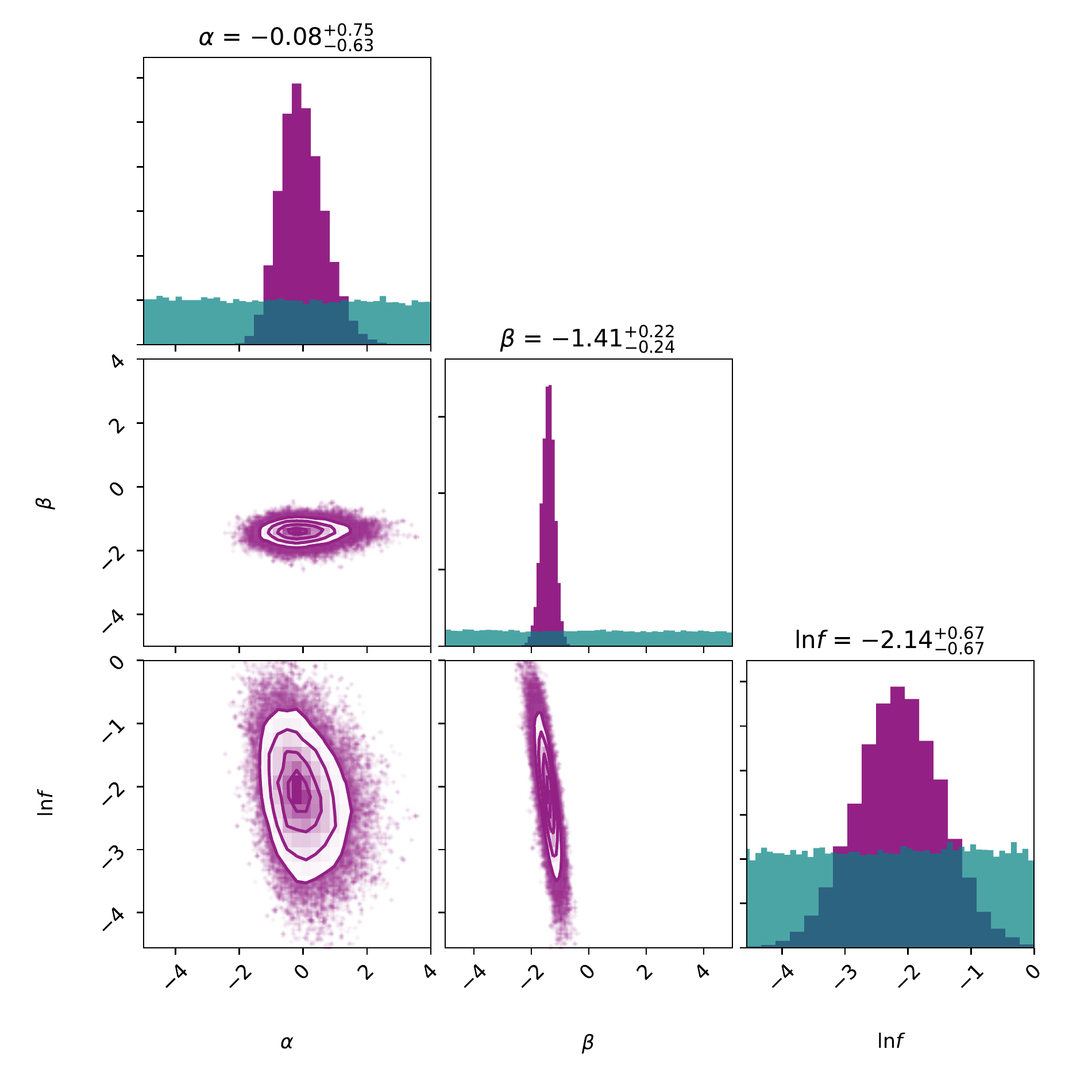}
\caption{\label{triangle_all} Results of the MCMC simulations for all the stars of the sample. The histogram represent the marginalized posterior probability distributions for our three parameters : $\alpha$, $\beta$ and $f$. Correlations plots for the parameters are also shown, with the solid contour lines corresponding to regions containing 68\%, 95\% and 99\% of the posterior. We use 200 walkers with 1000 steps.  Here, purples are highest and whites are lowest values of the likelihoods. The posterior distributions of the priors are also shown in teal in the histograms, for reference. }
 \end{figure*}   

\subsection{Constraining the distribution of companions}

We used a Markov Chain Monte Carlo approach to constrain the distribution of companions as a function of their mass and semi-major axis. We used the same mass--semi-major axis grid in the 5--5000~au interval for the calculation of the completeness maps, and we use index $i$ to refer to a given bin in mass, index $j$ to refer to a given bin in semi-major axis, index $s$ to refer to a given bin in projected separation axis and index $k$ to refer to a given star out of the total sample of $N$ stars surveyed. The completeness maps as a function of the projected separation, calculated earlier, are noted as $C_{k,is}$. The set $\{d_{k,is}\}$ denotes the detection made by the observations, such that $d_{k,is}$ is 1 if there is a planet detected in bin $i,s$ for star $k$, otherwise $d_{k,is}$ is 0. In our calculations, detected companions were assigned to the projected separation bin $s$ when they were detected in the images used in our study.

As a first case, we assumed that the distribution of planets follows a power law in mass and semi-major axis,
\begin{equation}\label{dn1}
{\rm d}^2n=f\dfrac{\alpha\beta}{C}M^{\alpha} a^{\beta} {\rm d}  M {\rm d} a\,,
\end{equation}
where d$n$ is the expected number of companions with a mass in the range $[M, M+$d$M]$ and semi-major axis in the range $[a, a+$d$a]$, $f$ is the overall mean number of planets per star (what we also call the frequency of planets), and $C$ is a normalization constant ensuring that the overall expected number of companions per star found by integrating d$n$ over the full mass and semi-major axis is equal to $f$.

For a given bin $(i,j)$, the expected number of companions is found by integrating over the bin, which yields,\\
if $\alpha \neq -1$ and $\beta \neq -1$ :
\begin{equation}\label{n_b}
n_{i,j}=\dfrac{f}{D}
\left[M_{i+1}^{\alpha+1} - M_{i}^{\alpha+1}\right]
\left[a_{i+1}^{\beta+1} - a_{i}^{\beta+1}\right]\, ,
\end{equation}
or if $\alpha = -1$ and $\beta = -1$
\begin{equation}\label{n_b}
n_{i,j}=\dfrac{f}{D} 
\ln \left(\dfrac{M_{i+1}}{M_{i}} \right)
\ln \left(\dfrac{a_{i+1}}{a_{i}} \right)\, ,
\end{equation}
or if $\alpha \neq -1$ and $\beta = -1$
\begin{equation}\label{n_b}
n_{i,j}=\dfrac{f}{D}
\left[M_{i+1}^\alpha - M_{i}^\alpha\right]
\ln \left(\dfrac{a_{i+1}}{a_{i}} \right)\, ,
\end{equation}
or if $\alpha = -1$ and $\beta \neq-1$
\begin{equation}\label{n_b}
n_{i,j}=\dfrac{f}{D}
\ln \left(\dfrac{M_{i+1}}{M_{i}} \right)
\left[a_{i+1}^\beta - a_{i}^\beta\right]\,, 
\end{equation}
where $D=(M_{max}^{\alpha+1}-M_{min}^{\alpha+1})(a_{max}^{\beta+1}-a_{min}^{\beta+1})$.

The number of companions expected in a bin $(i,s)$, of given mass and projected separation, is equal to the number of companions expected in bin $(i,j)$, of mass and semi-major axis, multiplied by the probability $p(s|j)$ of the companions to be observed at projected separation $s$ given their semi-major axis $j$, and summed up over all semi-major axis bins,
\begin{equation}
n_{is}=\sum_j n_{ij}p(s|j) \,.
\end{equation}
The probability $p(s|j)$ is computed using a Monte Carlo simulation assuming the same eccentricity distribution as before, and accounting for random orientations and phases of the observations.
Factoring in the completeness of the observations, the expected number of {\em detected} planets in bin $(i,s)$ for star $k$ is thus $C_{k,is}n_{is}$. Assuming that the presence of a planet in a given bin does not depend on the presence of other planets in other bins and using Poisson statistics for each bin and each star, we have that the probability $P$ of obtaining the observed results in a given bin given the assumed models is given by :
\begin{equation}
P(\{d_{k,is}\}|\{\alpha,\beta,f\})=e^{-C_{k,is}n_{is}}(C_{k,is}n_{is})^{d_{k,is}}\,.
\end{equation}

Thus, the likelihood of the whole survey results is obtained by multiplying the above probability for all bins and all stars :
\begin{equation}
\label{like}
\mathcal{L}(\{d_{k,is}\}|\{\alpha,\beta,f\})=\prod_{k=1}^{N} \prod_{i,s} e^{-C_{k,is}n_{is}}(C_{k,is}n_{is})^{d_{k,is}}\,,
\end{equation}
or 
\begin{equation}\label{loglike}
\ln{\mathcal{L}}=\sum_{i,j,k} \ln{[(C_{k,is}n_{is})^{d_{k,is}}]}-C_{k,is}n_{is}\,.
\end{equation}
This is the form we used in the calculations that follow. The set $\{d_{k,is}\}$ for our survey includes the detection of seven companions, as mentioned in Section \ref{sample}. In this section, we consider the full range of separations from 5 to 5000 au, rather than only 20 to 5000 au as in the previous analysis; a justification will be provided later.

To constrain the parameters $\alpha$, $\beta$ and $f$ that define the companion distribution, we used  the $emcee$ \citep{foreman-mackey_emcee:_2013} Python implementation of the affine-invariant MCMC ensemble sampler of \cite{goodman_ensemble_2010}. The MCMC sampler iteratively generates, for each of several random walkers, a sequence of samples for the three parameters in our model. We used uniform priors on all parameters (in log scale for $f$) and we defined the starting parameter values for the walkers to be drawn randomly from a uniform distribution between $-3$ and $1$ for $\log{f}$, between $-4.9$ and $4.9$ for $\alpha$, and between $-4.9$ and $4.9$ for $\beta$. We discarded the first 25\% of the steps as the burn-in phase and considered that remaining of the samples was representative of the posterior densities. The likelihood function is computed at each iteration, for each set of parameters. At each step, the sampler tries to move the walkers randomly in the parameter space: if the new set of parameters corresponds to a higher probability density part of the posterior distribution, then the move is accepted, otherwise, the new set can be accepted or rejected depending on the trial positions. The sampler thus mostly probes higher probability region of the parameter space and the final output samples are representative of the posterior distributions for each parameter of the model.

Figure~\ref{triangle_all} shows the results for 200 walkers and 1000 steps. The results indicate that $\alpha=-0.08^{+0.75}_{-0.63}$, $\beta=-1.41^{+0.22}_{-0.24}$ and $f =0.12^{+0.11}_{-0.06}$, where the uncertainty corresponds to 68\% confidence intervals. This indicates an increased planet occurence for smaller semi-major axes, while the planet mass distribution shows a marginal decrease with mass. The parameters $\alpha$ and $\beta$ show no correlation between each others.

Results from RV surveys have shown that the host star mass is correlated with the presence of planets \citep{johnson_giant_2010}. In the case of planets on wide orbits, there seems to be no significant trend in planet frequency with host mass \citep{bowler_imaging_2016} or a moderate trend \citep{lannier_massive:_2016} that would indicate that planets on wide orbits may be more common around more massive stars. To investigate this, we added a dependence on the host star mass in the distribution of planets. The planet distribution then becomes, 
\begin{equation}\label{dn_b2}
{\rm d}^2n=f\dfrac{\alpha\beta}{C}
\left(\dfrac{M_\star}{M_{\odot}}\right)^\gamma M^\alpha a^\beta \mathrm{d} M \mathrm{d} a\,.
\end{equation}

The mass of each star in the sample was estimated from either its spectral type or its $J$-band absolute magnitude. For stars with spectral types from late-B to late-K, we used the evolution models of \cite{siess_internet_2000} to estimate the mass from the spectral type and the age. For stars with spectral types of M0 or later, we used models from \cite{baraffe_new_2015} to estimate the mass from the $J$-band magnitude and the age. The masses for the earlier-type stars ($<$late-B) were taken from \cite{lafreniere_adaptive_2014}, where they were estimated using the evolution models of \cite{schaller_new_1992}. Lastly, the mass of HIP~100751 was taken from \cite{david_ages_2015}. 

\begin{figure*}[h] 
\centering
\includegraphics[width=\linewidth]{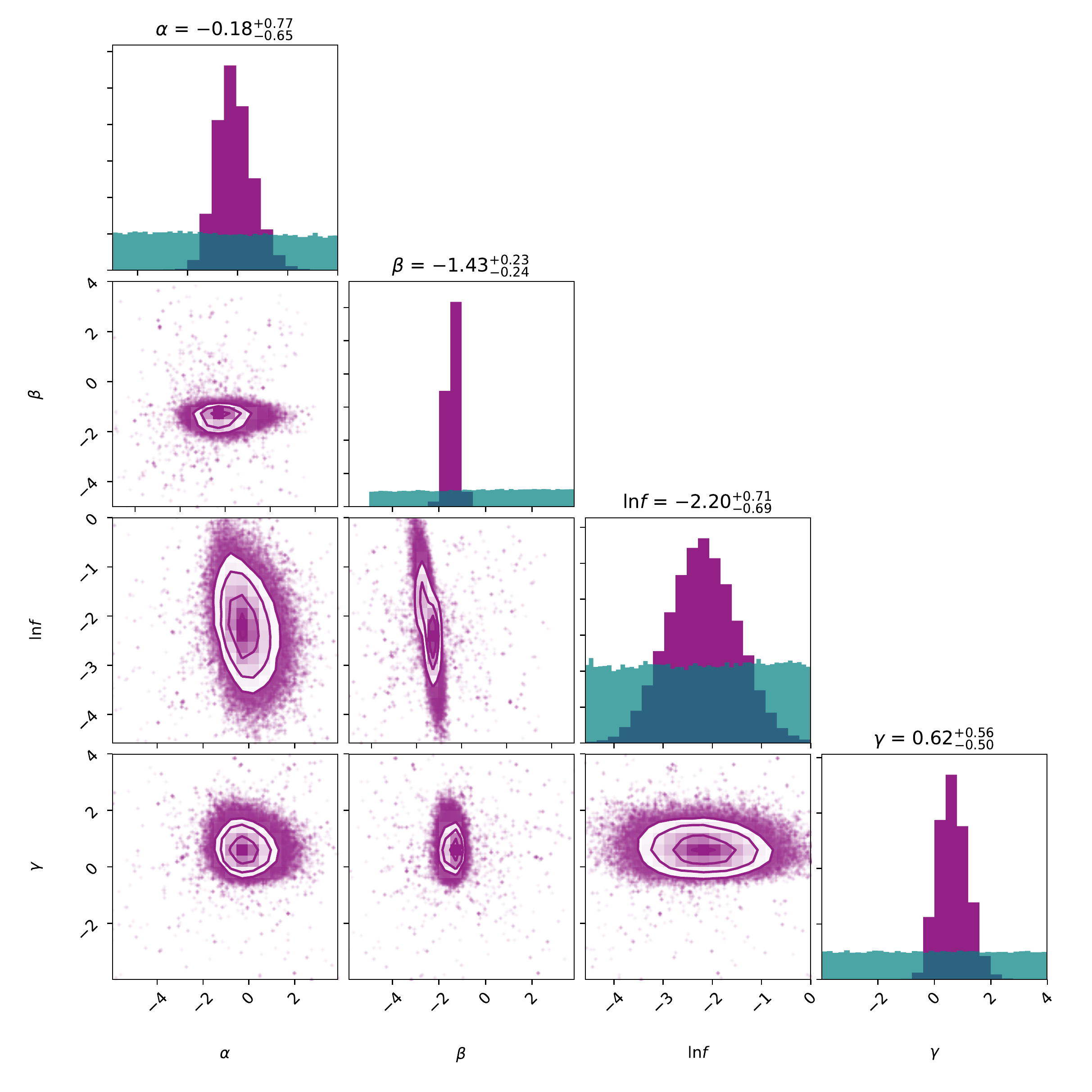}
\caption{\label{triangle_all_4} Same as Figure~\ref{triangle_all} but with 4 parameters and using distribution of planets from Equation~\ref{dn_b2}.
}
 \end{figure*}

This new model has four parameters. In our MCMC, we used the same initial ranges for our walkers for the three parameters we had previously, and for $\gamma$ we used random values in the range $-4.9$ to $4.9$. We used uniform priors on all parameters, 200 walkers and 1000 steps. The results are shown on Figure~\ref{triangle_all_4}. No correlation is seen between $\alpha$ and $\beta$, $\gamma$ and $\beta$ or $\alpha$ and $\gamma$. However, the frequency $f$ is correlated to all other parameters. Our results indicate that the best parameter values are $\alpha=-0.18^{+0.77}_{-0.65}$, $\beta=-1.43^{+0.23}_{-0.24}$,  $\gamma=0.62^{+0.56}_{-0.50}$ and $f=0.11^{+0.11}_{-0.05}$. The values for $\alpha$, $\beta$ and $f$ are consistent within uncertainties to the values obtained with the previous models. The added parameter $\gamma$ shows that the number of planets is correlated with the host star mass, such that massive stars host more planets in the separation and mass domains considered here.

As mentioned above, in this section we considered the full range from 5--5000~au instead of the 20--5000~range used in section~\ref{freq3.1}. To verify our choice of orbital separation range, we repeated the calculations in this section but over the 20--5000 au range, and in the case where the distribution is described by Equation~\ref{dn1}, we obtained $\alpha=-0.10^{+0.75}_{-0.65}$, $\beta=-1.58^{+0.29}_{-0.26}$ and $f =0.23^{+0.35}_{-0.15}$. Thus, both semi-major axis intervals favor similar alpha and beta, but the overall planet frequency is significantly higher and has much larger uncertainties for the 20-5000 AU interval  (although both agree within uncertainties). The higher uncertainty on the frequency for the 20--5000~au range can be understood on the basis of the favored slope of the semi-major axes distribution, which puts much more planets on shorter orbits. The effect of a change in planet frequency is thus more pronounced at the shortest separations, and neglecting the observational information that we have in the 5-20~au interval, even if incomplete, has a big impact on the frequency uncertainty. For the analyses in this section, we decided to keep the range that provides the lowest uncertainties, namely the 5—5000AU interval.

\subsection{Comparison with cold-start models}
The analysis described in the earlier sections uses hot-start models. However, it is possible that planets on wide orbits formed through a cold-start. For planets at more than hundreds of au separations, this would likely mean that they formed in the disk at smaller separations and migrated out. In the cold-start models, an accretion shock is created by free-falling gas onto the protoplanet, which irradiates the gravitational potential energy away from the core. This leaves newly-formed planets with low entropies and luminosities. Young massive planets are much fainter in cold-start models than in hot-start models. This effect is particularly important for young objects as the luminosity for both hot and cold-start models is similar at 200\,Myr and beyond, as the initial conditions effects are washed away by evolution.

To investigate the impact that a cold-start formation would have on our results, we did one more analysis similar to those presented in Section~\ref{freq3.1} but this time we used cold-start models from \cite{fortney_synthetic_2008}. 
The \cite{fortney_synthetic_2008} models give $T_{\rm eff}$ and $R$ for given masses as a function of age. To use these models for our purpose, we first had to interpolate the given values at the ages of the stars in our sample and on our grid of masses; because the models were available only up to 10~\mj, we linearly extrapolated the models from 10 to 20~\mj\ to complete our grid, neglecting luminosity bursts due to deuterium burning for objects more massive than 13~\mj, and we calculated synthetic magnitudes for all filters used in our study. To do so, we used synthetic spectra from the BT-Settl atmospheric models, scaled to the luminosity of the models, in combination with the appropriate filter response functions.
The synthetic spectra are only available for $\log{g} =4$  and  effective temperatures ranging from 400 to 600~K; when the surface gravity of the models was below 4, we used a synthetic spectrum with $\log{g}=4$. Furthermore, the cold-start models yield planets with effective temperatures in the range 170--560\,K, extending much below the lowest temperature (400\,K) of the synthetic spectra. The effective temperatures of the cold-start models in the range 400--600\,K were thus interpolated into the atmospheric grid at temperatures of 400, 500 or 600\,K, effective temperatures in 350--400\,K were extrapolated, while temperatures below 350\,K were considered too cold to be detected. We were then able to calculate cold-start contrast maps for each star. The completeness maps for the cold-start models are not as good as for the hot-start models, since companions of mass 1--20~\mj\ are much fainter in the former models. Indeed, a 10~\mj\ can be five magnitudes fainter in cold-start models than in hot-start models. Still, with the cold-start completeness maps, we inferred the frequency of companions of masses between 1 and 20~\mj, and separations of 5--5000~au as we did previously in Section~\ref{freq3.1}. In the case of the cold-start models, the set $\{d_k\}$ is 0 for all targets, as all detected companions would be more massive than 20 \mj\ according to those models. We obtained an upper limit of 5.2\%, at 95\% confidence level, for companions with masses in 1--20~\mj\ and separation in the 5--5000~au range, which is only slightly higher than the companion frequency inferred from hot-start models.

\section{Discussion}\label{discussion}

The frequency of Jupiter-like planets has been evaluated many times before from surveys made with other techniques than direct imaging. One study often quoted is that by \cite{cumming_keck_2008} mentioned in the introduction. Based on 8 years of precise radial velocity measurements from the Keck Planet Search, they inferred that 10.5\% of solar type stars have a companion with a mass of 0.3--10~\mj\ and a semi-major axis below 3~au. Similarly, based on the results of the HARPS planet search, \cite{mayor_harps_2011} found that 9.7$\pm$1.3\% of stars host a gas giant ($>$0.3~\mj) with a semi-major axis $<$4.6~au. Taken at face values and in comparison with the RV results, our results indicate that giant planets are less frequent above 5~au than below, even when summing the planet population all the way up to 5000~au. One possible caveat here is that our imaging survey has very little sensitivity to planets below 1--2~\mj, and thus that a population low-mass giant planets may be unnaccounted for in our results.

Estimates of giant planet occurrence were also derived from microlensing surveys. Based on the OGLE survey follow-up by the PLANET collaboration, \cite{cassan_one_2012} find that 17$^{+6}_{-9}$\% of stars host massive planets ($0.3-10$~\mj) with a semi-major axis between 0.5 and 10\,au. This frequency is marginally higher than the one we infer here from direct imaging surveys, 2.61$^{+6.97}_{-1.00}$\% for the 20--5000\,au range. If we assume that about 10\% of the microlensing survey results is accounted for by planets below 5\,au, per the above RV results, then the remainder would be in very good agreement with our results. In turn, this would indicate that within our range of sensitivity most of the planets at the larger separations would be located toward the small semi-major axes, which is indeed as we observed in our sample.

Another caveat to our results is that our mass determinations are indirect, relying on mostly uncalibrated evolution models. If young giant planets are much fainter than expected by hot-start models, then possibly much more than currently estimated could have been missed by the observations, leading to an underestimate of giant planet occurrence at large separations. Our results based on the cold-start models suggest that this is however not the case. Giant planets thus seem to be less frequent at large separations than small separations even when applying cold-start models.

It has often been assumed that the companion mass and semi-major axis distribution of the radial velocity planets can be extrapolated for planets onto larger orbits, at least up to some point. Converting the \cite{cumming_keck_2008} period distribution mentioned in the introduction into a semi-major axis power-law distribution, in line with eq.~\ref{dn1}, yields a value of $\beta=-0.69\pm0.15$. This means that giant planets on short orbits are more common than on wide orbits.
Our value for $\beta$ of -1.43$^{+0.23}_{-0.24}$, which is significantly different from that of \cite{cumming_keck_2008}, possibly hints that the more massive planets (several \mj\ or more) on orbits $\gg5$\,au probed here are part of a different population than the less massive RV planets at $\la 3$\,au. 

\begin{figure}[htbp]
	\centering
	\includegraphics[width=8cm]{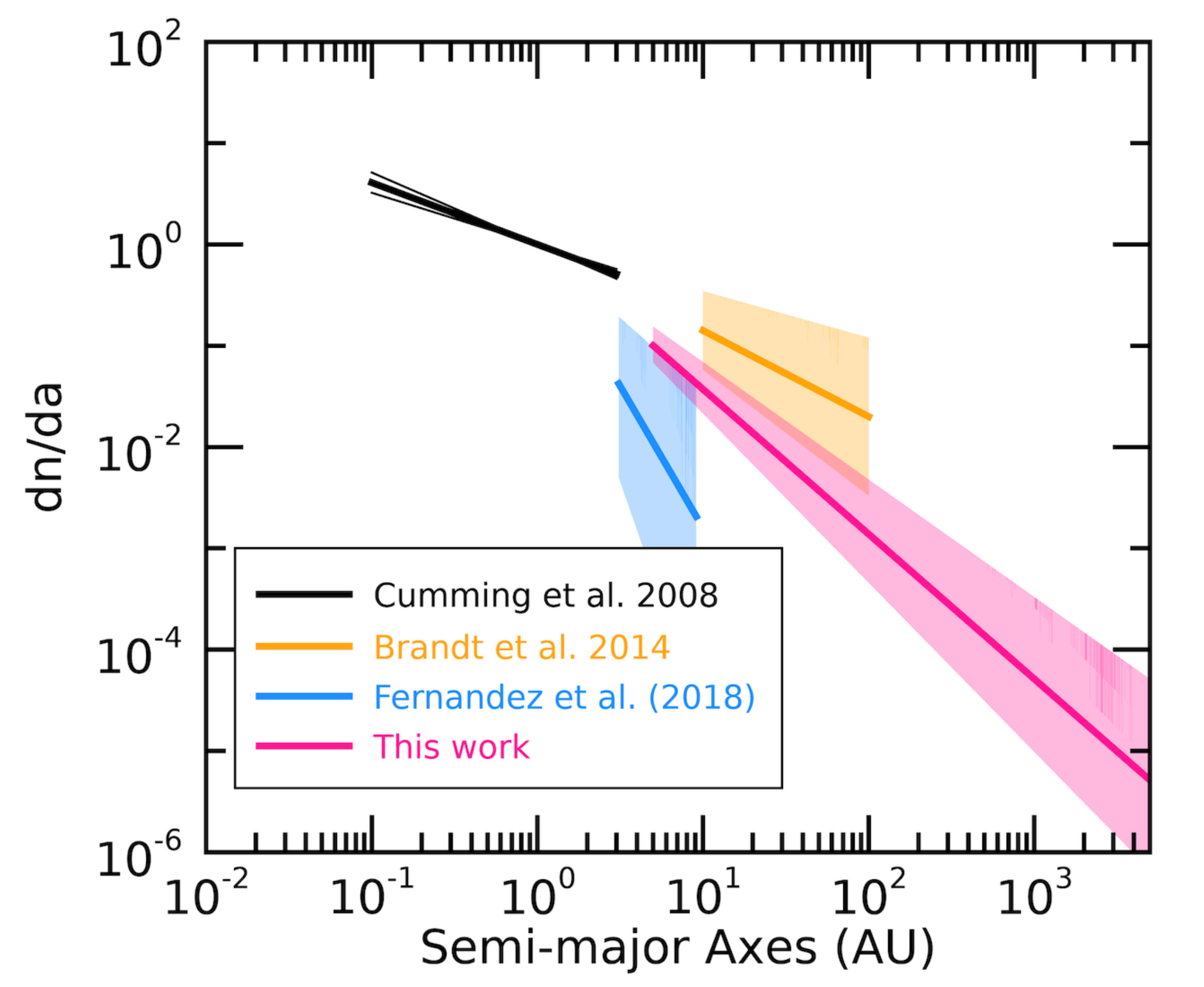}
	\caption{\label{distrsep} Comparison of the slopes of various power-law semi-major axis distributions of companions ($dn=a^\beta$).  The RV distribution of planets from \cite{cumming_keck_2008} is shown in black. The broken power law distribution of transiting and RV planets from \cite{fernandes_hints_2018} is drawn in blue; the turnover point is at 3\,au. The direct imaging distribution of brown dwarfs from \cite{brandt_statistical_2014} is drawn in orange. Only the slopes are depicted here, and all curves are normalized at a semi-major axis of 1~au.}
 \end{figure}

Figure~\ref{distrsep} compares the slopes of various power-law distributions in planet semi-major axes found in the literature. The slopes obtained in this work are shown in pink and are compared with the slopes from RV planets distribution from \cite{cumming_keck_2008} in black and \cite{fernandes_hints_2018} (asymmetrical distribution) in blue. The direct imaging distribution of brown dwarfs from \cite{brandt_statistical_2014} is drawn in orange. While the slope from \cite{cumming_keck_2008} is not consistent with the one we measure, as noted earlier, the slope from \cite{fernandes_hints_2018} at separations greater than the snow line is marginally consistent. The slope reported by \cite{brandt_statistical_2014} for more massive, 5--70~\mj, companions is also marginally consistent with ours. Overall, we measure a sharp decrease with semi-major axis, which is broadly consistent with the distribution of planets with a semi-major axis greater than 3\,au seen with RV and with brown dwarfs companions to main-sequence stars.

\begin{figure}[htbp]
	\centering
	\includegraphics[width=8cm]{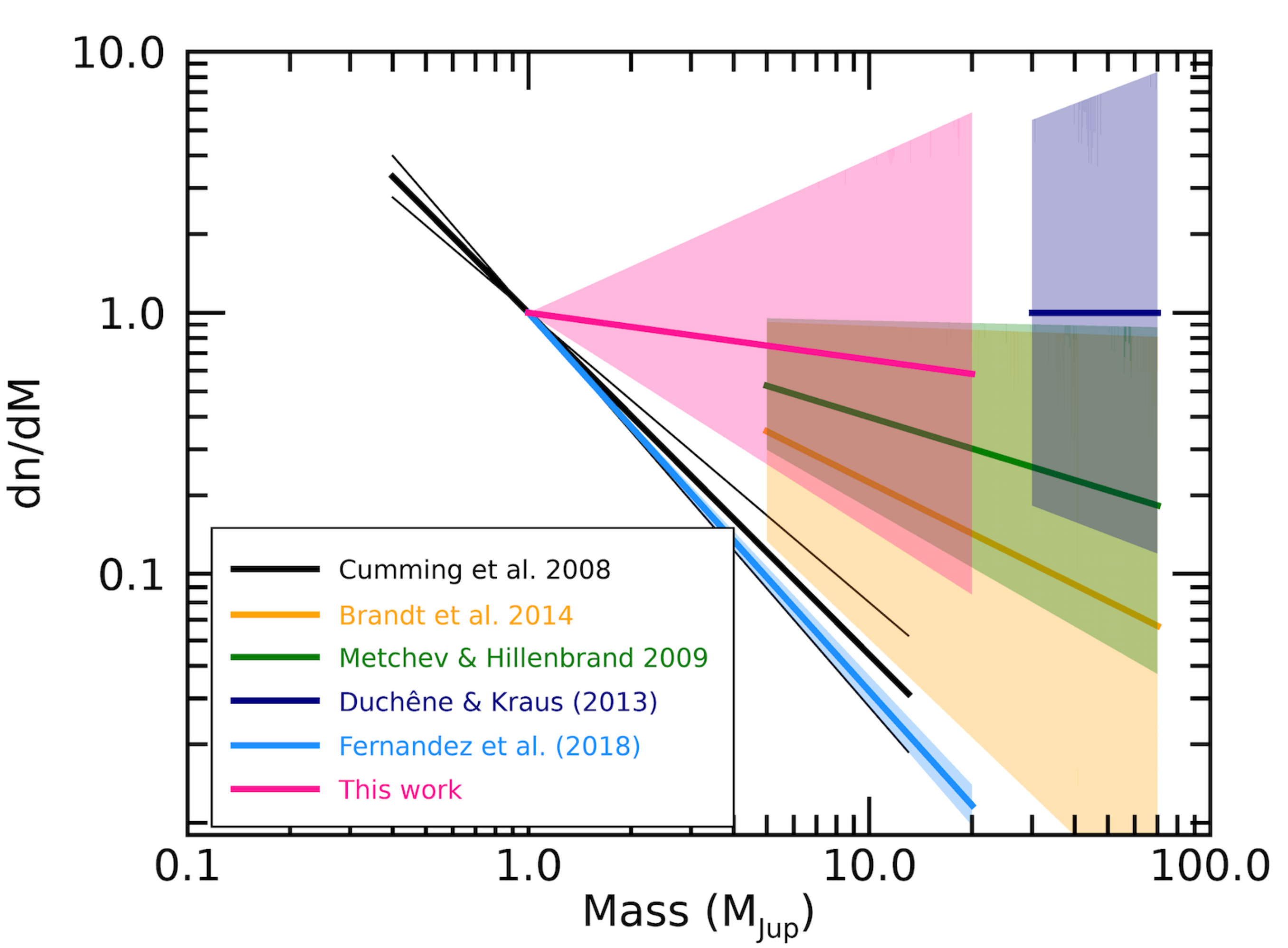}
 \caption{\label{distrmass} Comparison of the slopes of various power-law mass distributions of companions ($dn=M^\alpha$). The RV distribution of planets from \cite{cumming_keck_2008} is shown in black and in blue for \cite{fernandes_hints_2018}. The direct imaging distribution of brown dwarfs from \cite{metchev_palomar/keck_2009} is drawn in green and in orange from \cite{brandt_statistical_2014}. The distribution of stellar companions from \cite{duchene_stellar_2013} is shown in navy blue. The distribution of companions from this work is drawn in pink. The slope our distribution is consistent with slopes from the distribution of brown dwarfs or stellar companions. Only the slopes are depicted here, and all curves are normalized at a mass of 1~\mj.
}
 \end{figure}

Figure~\ref{distrmass} compare the slopes of various power-law distributions in mass. Our results, shown in pink, are in good agreement with the slope of the mass distribution of brown dwarf companions from \cite{metchev_palomar/keck_2009} in green and \cite{brandt_statistical_2014} in orange. They also agree with the slope for stellar companions at larger separations from \cite{duchene_stellar_2013}, shown in navy blue. However, it is not in agreement with the slopes of the distributions of RV planets from \cite{cumming_keck_2008} in black and \cite{fernandes_hints_2018} in blue. This may suggest that our survey is probing the low-mass tail of the brown dwarfs and stellar companion distribution rather than the continuation of the distribution of planets observed at smaller semi-major axes.

\cite{vigan_vlt/naco_2017} compiled 12 direct imaging surveys and compared the results to models based on the gravitational instability formation scenario from \cite{forgan_towards_2013}. In that study they showed that, assuming that companions form by the gravitational instability process, the models predict that the occurrence of companions increases with separation between 1 and 20\,au but decreases slowly with separation beyond. This change in the distribution with semi-major axis might be an evidence for a change of populations where the closer planets would be a population of non-scattered planets while the planets on wide orbits would be coming from a population of scattered planets. Qualitatively, our results at large separations agree with the simulations as we find a number of companions that decreases with semi-major axes. However, our slope is steeper that the slopes presented in \cite{vigan_vlt/naco_2017}.

Surveys focusing on probing the binary fraction of stars of all spectral types tend to show that the binaries with a low-mass component decline in number and have closer separations \citep{raghavan_survey_2010}. Also, binary fraction decrease with decreasing mass \citep{chabrier_review_2005,fontanive_constraining_2018}. Those results are consistent with our work, as our results show that companions are more frequent for more massive host stars.

\section{Conclusions}\label{conclu}
We used an MCMC analysis to put constraints on the distribution of 1--20\,\mj\ companions at separations of 5--5000\,au from a compilation of direct imaging surveys using the DIVA archives, a survey of Upper Scorpius, the PSYM-WIDE survey and the WEIRD survey. We used a distribution of planets in the form of a power-law in mass and semi-major axis of planets and host star mass. In general, we found that the occurrence of planets increases with smaller planet masses, closer orbits, and around more massive stars. Moreover, our constraints on the mass distribution shows that it is in better agreement with the mass distribution of brown dwarfs and stellar companions than it is with the distribution of planetary companions found by radial velocity at smaller separations.

The constraints on the distribution of companions found in this work depend on and are limited by the number of planets that the sample holds. In particular, while a wide range of semi-major axis is covered by the seven planets, the range in mass is rather narrow, as all the planets have very similar masses. This prevents us from reaching strong constraints on the $\alpha$ parameter controlling the power-law distribution in planet mass. Thus, the search for planets using direct imaging should continue, to uncover a larger and more diverse sample of planets enabling to better constrain their distribution.

In this work, we have chosen to fit our sample with a single power-law distribution. The next step in this project would be to use different distributions, for example a broken power-law in planet mass. This particular distribution would be motivated for instance by the work from \cite{santos_observational_2017} who have found evidence of a change in the population of giant planets at 4~\mj. 
They suggest that the lower-mass planets are formed by the core accretion process, while the more massive planets are mainly formed through the gravitational instability scenario, with an overlap of the two processes at 4~\mj. Also, \cite{reggiani_vlt/naco_2016}  suggested a superposition of two different populations to explain their null results from direct imaging. They coupled the brown dwarf companion distribution to the planet companion distribution truncated at about 100\,au. This new distribution has a minimum for companion masses in 10--50~\mj, which can explain the lack of objects in this range of masses without having to introduce another formation process for brown dwarfs.  This is another distribution that could be investigated with our sample, or preferably, with an expanded sample containing more detected companions and spanning a wider range of companion masses.

\acknowledgments
The authors thank the anonymous referee for constructive comments and suggestions that improved the overall quality of the paper.

This work was financially supported by the Natural Sciences and Engineering Research Council (NSERC) of Canada and the Fond de Recherche Québécois—Nature et Technologie (FRQNT; Québec). 

This publication makes use of data products
from the Two Micron All Sky Survey, which is a joint project of the University of Massachusetts and the Infrared Processing and Analysis Center, and funded by the National Aeronautics and Space Administration and the National Science Foundation, of the NASA Astrophysics Data System Bibliographic Services, the VizieR catalog access tool, and the SIMBAD database operated at CDS, Strasbourg, France.

This research has made use of the Direct Imaging Virtual Archive (DIVA), operated at CeSAM/LAM, Marseille, France.

This work has made use of data from the European Space Agency (ESA) mission {\it Gaia} (\url{https://www.cosmos.esa.int/gaia}), processed by
the {\it Gaia} Data Processing and Analysis Consortium (DPAC,
\url{https://www.cosmos.esa.int/web/gaia/dpac/consortium}). Funding
for the DPAC has been provided by national institutions, in particular
the institutions participating in the {\it Gaia} Multilateral Agreement.
\software{emcee}

\startlongtable
\begin{deluxetable*}{cccccccccccc}
\tablecaption{Properties of the sample of bona fide members\label{listecomplete}}
\tablewidth{700pt}
\tabletypesize{\scriptsize}
\tablehead{
\colhead{Name} & \colhead{RA} &
\colhead{DEC} & \colhead{SpT} &
\colhead{$\mu_{\alpha}cos\delta$} & \colhead{$\mu_{\delta}$} &
\colhead{Association} & \colhead{Distance} & \colhead{Masses} & \colhead{Surveys} \\
\colhead{} & \colhead{(J2000.0)} & \colhead{(J2000.0)} & \colhead{} &
\colhead{(mas/yr)} & \colhead{(mas/yr)} &  \colhead{} &\colhead{(pc)} &
\colhead{(M$_\odot$)}
}
\startdata
                  HIP 490   &                00 05 52.54   &                -41 45 11.0   &                        G0V   &             97.53$\pm$0.38   &            -76.27$\pm$0.44   &                        THA   &           39.385$\pm$0.915   &                       1.26   &                          3  \\
                   HD 203   &                00 06 50.08   &                -23 06 27.1   &                      F2 IV   &             96.77$\pm$0.13   &            -47.12$\pm$0.06   &                       BPIC   &           39.960$\pm$0.099   &                       1.53   &                          3  \\
                 HIP 1113   &                00 13 53.01   &                -74 41 17.8   &                        G8V   &             83.53$\pm$0.78   &            -47.89$\pm$0.75   &                        THA   &           44.404$\pm$1.616   &                       1.10   &                          3  \\
                   HD 984   &                00 14 10.25   &                -07 11 56.8   &                       F5 V   &            104.53$\pm$0.15   &            -67.91$\pm$0.06   &                        COL   &           45.911$\pm$0.118   &                       1.40   &                          3  \\
  2MASS J00172353-6645124   &                00 17 23.53   &                -66 45 12.4   &                       M2.5   &            102.90$\pm$1.00   &            -15.00$\pm$1.00   &                        ABD   &           51.241$\pm$0.120   &                       0.74   &                         10  \\
                V* PW And   &                00 18 20.89   &                 30 57 22.1   &                       K0Ve   &            143.19$\pm$0.09   &           -171.11$\pm$0.06   &                        ABD   &           29.450$\pm$0.133   &                       1.00   &                         12  \\
\enddata
\end{deluxetable*}  

\bibliographystyle{yahapj}
\bibliography{bib}

\end{document}